\documentstyle[aps,epsf]{revtex}

\def\calS{{\cal S}}
\def\Tr{{\rm Tr}}

\begin{document}

\title{The Liquid-Gas Phase Transitions in a Multicomponent Nuclear System
with Coulomb and Surface Effects}

\author{S.J. Lee\cite{byline}
and
A.Z. Mekjian
}

\address{Department of Physics and Astronomy, Rutgers University,
Piscataway, New Jersey 08855} 

%\date{\today}  June 26, 2000  PRC

\maketitle

\begin{abstract}
The liquid-gas phase transition is studied in a multi-component nuclear
system using a local Skyrme interaction with Coulomb and surface effects.
Some features are qualitatively the same as the results of
M\"{u}ller and Serot which uses relativistic mean field
without Coulomb and surface effects.
Surface tension brings the coexistance binodal surface to lower pressure.
The Coulomb interaction makes the binodal surface smaller
and cause another pair of binodal points at low pressure
and large proton fraction
with less protons in liquid phase and more protons in gas phase.
\end{abstract}

%\pacs

%\narrowtext

\section{Introduction}

The liquid-gas phase transition in nuclei was first studied using 
a Skyrme interaction and focussed mostly on one component systems
of just nucleons even though expressions were developed for
two component systems of protons and neutrons \cite{skyrmp1}.
The phase transition aspects are considerably easier to study
in one component systems rather than two, where for example
one has to deal with separate proton and neutron chemical potentials
for charge and nucleon number conservation.
Initial fragmentation models also mostly dealt with one
component systems.
Such fragmentation studies gave the first evidence for the
liquid-gas phase transition in nuclear systems \cite{finn,hirsch}.
Since then, the liquid-gas phase transition has been extensively 
studied experimentally and theoretically.
Several reviews exist on this topic \cite{csern,lynch,dasgup}.

Because of the two component nature of real nuclear systems,
an analysis of liquid-gas phase transition in these systems is important.
Some preliminary results were reported in Ref.\cite{skyrmp1},
and a very detailed study was done by M\"{u}ller and Serot \cite{serot}
who used a relativistic mean field model to develop the main
thermodynamic properties of asymmetric nuclear matter.
One interesting new aspect of two component system compared to
one component system is that the phase transition is a second order
transition in their approach.
The importance of the number of components on the order of the
transition was pointed out by Glendenning \cite{glend}.
Another interesting aspect of two component systems is the possibility
of having different proton--neutron ratios in the liquid and gas
phases because of the symmetry energy.
The study of nuclear systems with arbitrary proton--neutron ratios
is important for radioactive beam experiments and in astrophysical
situations as in neutron stars.
Because of the extra degree of freedom associated with varying
proton fraction $y$ in the two phases,
the phase diagram has a higher dimensionality and now becomes
a surface in pressure $P$, temperature $T$, and proton fraction $y$
or nucleon density $\rho$.
For one component systems the phase diagram is represented
as a binodal curve of $P$ versus density $\rho$ or volume $V$,
whose end points at fixed $T$ give the liquid and gas densities.
The Maxwell pressure versus $T$ in one component systems is a line
that terminate at the critical temperature $T_c$.
For two component systems, the binodal surface associated with
phase coexistance in $(P, T, y)$
now contains a line of critical points for different values of $y$,
a line for the Maxwell pressure versus $T$ for a symmetric system
at $y = 0.5$ which  has a the smallest $P$ and is the same as for a 
one component system, and a line of maximal asymmetry.
At fixed $T$, $P$ versus $y$ of binodal points form loops
and $(dy/dP)_T = 0$ gives a point of maximal asymmetry or
smallest proton ratio on the binodal surface.

In this paper we extend the initial study of Ref.\cite{skyrmp1}
using Skyrme interaction in a similar way as done in Ref.\cite{serot}.
Some features are qualitatively the same as in Ref.\cite{serot},
but quantitatively differ because of the different interaction.
Our equation of state based on the Skyrme interaction 
with Coulomb and surface effects
has some feature also not present in Ref.\cite{serot}
which will also be discussed.
According to the results of Ref.\cite{serot} which has no Coulomb interaction, 
the liquid phase has higher proton fraction than the gas phase
of mixed phases in the coexistent state.
A real nucleus has less protons than neutrons due to Coulomb interaction.
For a given $A = N + Z$ stability is determined by Coulomb
and symmetry energy effects.
Since a stable finite nucleus has zero internal pressure
while the gas phase having positive pressure \cite{prc56},
we also need to consider surface effects.
In this paper we consider the effects of Coulomb interaction 
and surface tension by considering uniform spherical finite nuclear system.

In Section II, the main equations for the thermodynamic properties of
hot nuclear matter in mean field theory are developed
as a function of density $\rho$, temperature $T$,
and proton fraction $y$ or neutron fraction $(1-y)$.
These include the pressure and chemical potentials, 
both neutron $\mu_n$ and proton $\mu_p$ where
phase equilibrium requires equality of the proton chemical potential
between the two phases and similarly equality of the
neutron chemical potentials at a given temperature and $P$.
At fixed $T$ and $P$, $\rho$ and $y$ are not independent
and we use this connection to simplify the analysis.
Section III contains the results of calculations performed using
the equations developed in Section II.
Conclusions are given in Section IV.

\section{Phase Transition in Mean Field Theory}

For phase transition we look at the pressure $P$ and
chemical potentials $\mu_q$ for each component of multi-component
system (neutron and proton for example)
as functions of temperautre $T$ and the densities $\rho_q$ of
the constituents.
These quantities can 
be obtained once we know the total energy functional $E$
as a function of the densities $\rho_q$ at a given temperature $T$.

At a given temperature $T = 1/\beta$,
the constituents are distributed in phase space according to
the Wigner function $f$ as
\begin{eqnarray}
 f(\vec r, \vec p) = \sum_q f_q(\vec r, \vec p) ,   \hspace{1.5cm}
 f_q(\vec r, \vec p) = \frac{\gamma}{h^3} \tilde f_q(\vec r, \vec p)
    = \frac{\gamma}{h^3} \frac{1}{e^{\beta(\epsilon_q-\mu_q)} + 1} 
            \label{wigf}
\end{eqnarray}
where the spin degneracy $\gamma = 2$ and 
$\epsilon_q$ and $\mu_q$ are the single particle energy
and the chemical potential of particle type $q$.
Then the particle density $\rho$ becomes
\begin{eqnarray}
 \rho(\vec r) = \sum_q \rho_q(\vec r) , &\hspace{1.5cm}&
 \rho_q(\vec r) = \int d^3 p f_q(\vec r, \vec p)  ,   \\
 A = \sum_q N_q = \int d^3 r \rho(\vec r) , &\hspace{1.5cm}&
 N_q = \int d^3 r \rho_q(\vec r)
     = \int d^3 r \int d^3 p f_q(\vec r, \vec p)
\end{eqnarray}
and the total energy is
\begin{eqnarray}
 E = \int d^3 r {\cal E}(\vec r)
   = \int d^3r \int d^3 p \frac{p^2}{2m} f(\vec r, \vec p)
    + \int d^3 r U(\vec r)
   = \int d^3 r \left[ {\cal E}_K(\vec r) + U(\vec r) \right] 
           \label{tenerg}
\end{eqnarray}
with the potential energy density $U(\vec r)$ and
the kinetic energy density ${\cal E}_K(\vec r)$.
These, in turn, give self-consistent equations for $\mu_q$ (or $p_{Fq}$)
in a mean field theory for fixed $T$ and $N_q$,
\begin{eqnarray}
 \epsilon_q &=& \frac{\delta E}{\delta f_q}
   = \frac{\delta {\cal E}(\vec r)}{\delta f_q(\vec r, \vec p)}
   = \frac{p^2}{2m} + \frac{\delta U}{\delta f_q}
   = \frac{p^2}{2m} + u_q(\vec r, \vec p)       \label{senerg} \\
 \mu_q &=& \left.\epsilon_q\right|_{p=p_{Fq}}
   = \frac{p_{Fq}^2}{2m} + u_q(\vec r, \vec p_{Fq})  \label{chempot}
\end{eqnarray}
where we define $p_{Fq}$ in Eq.(\ref{chempot}) and consider it
as an effective Fermi momentum at $T$ of particle $q$,
which is the momentum associated with the chemical potential,
and $u_q = \frac{\delta U}{\delta f_q}$ is the single particle
potential of particle $q$ which may be momentum dependent in general.

The pressure $P$ can be defined dynamically
from the total momentum conservation law,
$\frac{d}{dt} \left[\int d^3 r \int d^3 p ~\vec p f\right] 
   = - \int d^3 r \vec\nabla_r \cdot \buildrel\leftrightarrow\over\Pi = 0$,
using the Vlasov equation \cite{bertsch}
\begin{eqnarray}
 \frac{\partial f_q}{\partial t}
 + (\vec\nabla_p \epsilon_q) \cdot (\vec\nabla_r f_q)
 - (\vec\nabla_r \epsilon_q) \cdot (\vec\nabla_p f_q)
 = 0 
\end{eqnarray}
or more generally from hydrodynamic consideration of TDHF in
phase space \cite{prc42q}
\begin{eqnarray}
\vec\nabla_r \cdot \buildrel\leftrightarrow\over\Pi
 &=& - \frac{d}{dt} \left[ \int d^3 p \vec p \sum_q f_q(\vec r, \vec p) \right]
  = - \sum_q \int d^3 p \vec p \left(\frac{\partial f_q}{\partial t}\right)
         \nonumber \\
 &=& \sum_q \int d^3 p \vec p
       ~\vec\nabla_r\cdot\left[(\vec\nabla_p\epsilon_q)f_q\right]
     + \sum_q \int d^3 p \hat{p} \cdot (\vec\nabla_r\epsilon_q) f_q 
\end{eqnarray}
where $\hat p$ is the unit vector in the direction of $\vec p$.
Using
$(\vec\nabla_r\epsilon_q)f_q = \vec\nabla_r(\epsilon_q f_q)
   - \epsilon_q \vec\nabla_r f_q   
 = \vec\nabla_r(\epsilon_q f_q) - \vec\nabla_r {\cal E}$,
the dynamical pressure tensor $\Pi_{ij}$ becomes
\begin{eqnarray}
 \Pi_{ij} &=& \sum_q \int d^3 p p_i \left(\nabla_p^j \epsilon_q \right) f_q
 + \delta_{ij} \left[\int d^3 p \sum_q \epsilon_q f_q - {\cal E} \right]
           \nonumber \\
 &=& \sum_q \int d^3 p p_i 
          \nabla_p^j \left(\frac{\delta {\cal E}}{\delta f_q}\right) f_q
 + \delta_{ij} \left[
   \sum_q \int d^3 p \left(\frac{\delta {\cal E}}{\delta f_q}\right) f_q
       - {\cal E} \right]   \nonumber \\
 &=& \sum_q \int d^3 p p_i \left[\frac{p_j}{m}
     + \nabla_p^j \left(\frac{\delta U}{\delta f_q}\right) \right] f_q
 + \delta_{ij} \left[\sum_q \int d^3 p 
       \left(\frac{\delta U}{\delta f_q}\right) f_q
       - U \right]      \label{ppi}
\end{eqnarray}
For a momentum independent potential, this becomes
\begin{eqnarray}
 \Pi_{ij} &=& \sum_q \int d^3 p \frac{p_i p_j}{m} f_q
    + \delta_{ij} \left[ \sum_q \left(\frac{\delta U}{\delta\rho_q}\right)
             \rho_q  - U \right]
    = \int d^3 p \sum_q \frac{p_i p_j}{m} f_q
    + \delta_{ij} \sum_q \rho_q^2 \frac{\delta(U/\rho_q)}{\delta\rho_q}
\end{eqnarray}
The diagonal elements are
\begin{eqnarray}
 P = \Pi_{ii} = \sum_q \int d^3 p \frac{p_i^2}{m} f_q
     + \sum_q \frac{\delta U}{\delta\rho_q} \rho_q - U
   = P_K + \sum_q u_q\rho_q - U
   = P_K + P_P    \label{pres}
\end{eqnarray}
where $P_K = \int d^3 p \frac{p_i^2}{m} f$ 
and $P_P = \sum_q u_q \rho_q - U$ are the kinetic pressure
and interaction (potential) pressure respectively.
In equilibrium $P$ can also be obtained by minimizing the total energy 
as a function of volume holding the number of particles fixed:
\begin{eqnarray}
 P = \Pi_{ii} = - \frac{d (E/A)}{dV} = \rho^2 \frac{d({\cal E}/\rho)}{d\rho}
\end{eqnarray}
for a single component uniform system.

From the distribution $\tilde f_q$ of Eq.(\ref{wigf}),
the entropy $S$ can be obtained as,
\begin{eqnarray}
 S &=& \sum_q S_q = \int d^3 r \calS = \int d^3 r \sum_q \calS_q , \\
 \calS_q &=& - \frac{\gamma}{h^3} \int d^3 p 
     \left[\tilde f_q \ln \tilde f_q + (1-\tilde f_q) \ln (1-\tilde f_q)\right]
                \nonumber \\
   &=& \frac{\gamma}{h^3} \int d^3 p 
     \frac{1}{3} \vec p \cdot (\vec\nabla_p \tilde f_q)
        \ln\left[\frac{\tilde f_q}{1-\tilde f_q}\right]
                \nonumber \\
   &=& \frac{\gamma}{h^3} \int d^3 p 
     \frac{1}{3} \vec p \cdot (\vec\nabla_p \tilde f_q)
        \beta(\mu_q - \epsilon_q)
\end{eqnarray}
using
$\vec\nabla_p \cdot (\vec p g) = 3 g + \vec p \cdot \vec\nabla_p g$.
The $\calS_q$  becomes, by partial integration \cite{prc41prs},
\begin{eqnarray}
 \calS_q &=& \beta \int d^3 p \frac{1}{3}
        \vec p \cdot \vec\nabla_p \left[(\mu_q - \epsilon_q) f_q\right]
   - \beta \int d^3 p \frac{1}{3}
        \vec p \cdot \vec\nabla_p(\mu_q - \epsilon_q) f_q
         \nonumber \\
   &=& \beta \int d^3 p (\epsilon_q - \mu_q) f_q
   + \beta \int d^3 p \frac{\vec p\cdot\vec\nabla_p\epsilon_q}{3} f_q
         \nonumber \\
   &=& \beta \int d^3 p \epsilon_q f_q
   + \beta \int d^3 p 
         \frac{\vec p\cdot\vec\nabla_p\epsilon_q}{3} f_q
   - \beta \mu_q \int d^3 p f_q     \label{sqden}
\end{eqnarray}
In equilibrium thermodynamics,
the thermodynamic grand potential $\Omega$, the Helmholtz free energy $F$,
and the Gibb's free energy $G$ are 
\begin{eqnarray}
 \Omega &=& - \int d^3 r P
    = - \frac{1}{\beta} \ln \Tr e^{-\beta(\hat H - \sum_q \mu_q \hat N_q)}
    = F - G
    = E - TS - \sum_q \mu_q N_q      ,   \\ 
 F &=& \int d^3 r {\cal F}
    = - \frac{1}{\beta} \ln \Tr e^{-\beta \hat H}
    = E - TS = T \int E d\beta  ,     \\
 G &=& \int d^3 r {\cal G}
    = \frac{1}{\beta} \ln \Tr e^{\beta\sum_q \mu_q \hat N}
    = \sum_q \mu_q N_q
    = \int d^3 r \sum_q \mu_q \rho_q     
\end{eqnarray}
Comparing Eq.(\ref{sqden}) with Eq.(\ref{ppi})
for diagonal elements using
 $p_i \nabla_p^i = \frac{1}{3} \vec p \cdot \vec\nabla_p$
(isotropic condition in momentum, $p_i^2 = \frac{1}{3} p^2$),
\begin{eqnarray}
 T \calS &=& {\cal E} + P - \sum_q \mu_q \rho_q = {\cal E} - {\cal F}
       = {\cal E}_K + P_K - \sum_q (\mu_q - u_q) \rho_q    ,      \\ 
 {\cal F} &=& \sum_q \mu_q \rho_q - P
            = \sum_q (\mu_q - u_q) \rho_q - P_K + U  ,    \\
 {\cal G} &=& \sum_q {\cal G}_q = \sum_q \mu_q \rho_q
\end{eqnarray}
The entropy can also be found from $dQ = TdS = C dT$;
\begin{eqnarray}
 \Delta S = \int \frac{dQ}{T}
   = \int \frac{C dT}{T} = \int\frac{1}{T} \frac{dE}{dT} dT
   = \int \beta \frac{dE}{d\beta} d\beta = \beta E - \int E d\beta
   = \beta \left(E - F\right)
\end{eqnarray}
The pressure and the chemical potential are also related to 
the free energy $F$ as
\begin{eqnarray}
 & P = - \left. \frac{\partial F}{\partial V} \right|_{T,A}
     = - \left. \frac{\partial \Omega}{\partial V} \right|_{T,A}  ,   \\
 & \mu = \left. \frac{\partial F}{\partial A} \right|_{T,V}
\end{eqnarray}
The specific heat capacity is given by 
from the entropy per particle, $\calS/\rho$,
\begin{eqnarray}
 c_P = T \left(\frac{\partial \calS/\rho}{\partial T}\right)_P
         \hspace{0.7cm} {\rm or} \hspace{0.7cm}
 c_V = T \left(\frac{\partial \calS/\rho}{\partial T}\right)_V
                  \label{cpisob}
\end{eqnarray}
To study the caloric curve or the specific heat we look at
the enegy per particle ${\cal E}/\rho$ and
the entropy per particle $\calS/\rho$.

For multi-phase multi-component system, 
the phase transition of each component can occur at different conditions
such as temperature or pressure.
However in general we can treat all the different possible combinations
of phases of each component in
the multi-component system as different phases of the system.
(As an example suppose we have a system of particles of type p
and particles of type n, each having a 
liquid-gas phase transition but at different temperatures.
Then the system can be in one of the following four phases;
a phase of liquid p and liquid n, a phase of liquid p and gas n,
a phase of gas p and liquid n, or a phase of gas p and gas n.)
Then, to separate each phase of the multi-component system,
we can use the volume fraction $\lambda_i$ of $i$-phase of total
volume $V$ which depends on $T$, 
\begin{eqnarray}
 \lambda_i &=& V_i/V  \hspace{0.7cm} {\rm with}
       \hspace{0.7cm} \sum_i \lambda_i = 1   ,   \\
 \rho_q &=& \sum_i \lambda_i \rho_q^i   ,   \\
 {\cal O}(\rho_q, T) &=& \sum_i \lambda_i {\cal O}(T, \rho_q^i)
\end{eqnarray}
where ${\cal O}$ is any obserbable per unit volume.
Within the spinodal instability region, there is no equilibrated phase.
Two or multi phases can coexist when the pressure $P$ and each chemical
potential $\mu_q$ are all the same among these phases, i.e.,
\begin{eqnarray}
 P_i &=& P_j  ,   \\
 \mu_q^i &=& \mu_q^j
\end{eqnarray}
with different values of $\rho_q^i$ and $\rho_q^j$ for all $i$ and $j$.
At the critical point
\begin{eqnarray}
 \frac{\partial P}{\partial\rho_q} 
  = \frac{\partial^2 P}{\partial\rho_q^2} = 0   ,    \label{binodp} \\
 \frac{\partial\mu_q}{\partial\rho_q} 
  = \frac{\partial^2\mu_q}{\partial\rho_q^2} = 0    \label{binodmu}
\end{eqnarray}
Spinodal instability occurs when $\frac{\partial P}{\partial\rho_q}$
is negative.

Once the potential energy $U$ in Eq.(\ref{tenerg}) is known,
then the possibility of a phase transition of the system
can be studied  using Eqs.(\ref{wigf}) -- (\ref{cpisob}).
The potential energy $U$ determines $\epsilon_q$ and $\mu_q$ and
the potential energy part of $E$ and $P$.
Then for fixed $T$ and $N_q$, the Wigner function $f$ and $p_{Fq}$ are
determined and thus the kinetic terms of $E$, $\mu_q$, and $P$.
Using these results, the entropy $\calS$ and $c_P$ can be determined.
Relativistic mean field theory is used for the interaction in Ref.\cite{serot}.
The role of the Coulomb interaction and surface energy or the
finite size effect of a nucleus are neglected in Ref.\cite{serot}
but will be included in the approach developed here.
Their result shows that the neutron
evaporates first as the energy of the system increases leaving more charge
concentrated in liquid phase.
In this approach the coupled equations of nucleon and mesons lead to a
highly nonlinear system and thus
the equation of state can be obtained only through self-consistent
iterative way. These problems may be simplified by using a nonrelativistic
zero-range Skyrme interaction of
\begin{eqnarray}
 v_{12} = t_0 (1+x_0 P^\sigma) \delta(\vec r_1 - \vec r_2)
        + \frac{t_3}{6} (1+x_3 P^\sigma)
     \rho^\alpha(\frac{\vec r_1 + \vec r_2}{2}) \delta(\vec r_1 - \vec r_2) 
\end{eqnarray}
For a nuclear system of proton ($\rho_p$) and neutron ($\rho_n$),
this gives the local potential energy density as
\begin{eqnarray}
 U(\rho_q) &=& \frac{t_0}{2} (1 + \frac{x_0}{2}) \rho^2
              - \frac{t_0}{2} (\frac{1}{2} + x_0) \sum_q \rho_q^2
         + \frac{t_3}{12} (1 + \frac{x_3}{2}) \rho^{\alpha+2}
           - \frac{t_3}{12} (\frac{1}{2} + x_3) \rho^\alpha \sum_q \rho_q^2
         + C \rho^\beta \rho_p^2 + C_s \rho^\eta
    \label{potene}
\end{eqnarray}
Here $C \rho^\beta = \frac{4\pi}{5} e^2 R^2$ 
and $C_s \rho^\eta = \frac{4\pi R^2 \sigma(\rho)}{V}
  = \frac{(4\pi r_0^2 \sigma)}{V^{1/3}} \rho^{2/3}$
when we approximate the Coulomb and surface effects as
coming from a finite uniform sphere of radius $R = r_0 A^{1/3}$
with total charge $Z$ ($U_C = \frac{3}{5}\frac{e^2 Z^2}{R V}$).
The typical values for the force parameters are given in Table \ref{tabl1}.
For a symmetric nucleus, $N = Z$, $\rho_q = \rho/2$, 
and thus
\begin{eqnarray}
 U(\rho) = \frac{3}{8} t_0 \rho^2 + \frac{3}{48} t_3 \rho^{\alpha+2}
         + C \rho^\beta \rho_p^2 + C_s \rho^\eta
\end{eqnarray}
This potential enegy determines the interaction dependent terms
of ${\cal E}$, $P$, $\epsilon_q$, and $\mu_q$
which depend on densities without explicit $T$ dependence.

%\vspace{5ex}
%\begin{table}[tbh]
%%  \label{tabl1}
%\end{table}
%\vspace{5ex}

For a momentum independent potential energy as in Eq.(\ref{potene}),
$\epsilon_q - \mu_q = (p^2 - p_{Fq}^2)/(2m)$ is independent of
the potential and 
\begin{eqnarray}
 f_q(\vec r, \vec p) = \frac{\gamma}{h^3} \tilde f_q(\vec r, \vec p) ,
    \hspace{5ex}
 \tilde f_q(\vec r, \vec p) = \frac{1}{e^{\beta(p^2 - p_{Fq}^2)/(2m)} + 1}
\end{eqnarray}
Thus we can evaluate the kinetic terms in ${\cal E}$,
$P$, and $\mu_q$ which are functions of $T$ and $p_{Fq}$.
Defining Fermi integral $F_\alpha(\eta)$
\begin{eqnarray}
 F_\alpha(\eta_q) &=& \int_0^\infty \frac{x^\alpha}{1 + e^{x-\eta_q}} dx
     = \left(\frac{\lambda^2}{4\pi\hbar^2}\right)^{\alpha+1}
       \int_0^\infty \frac{2 p^{2\alpha+1} dp}
             {1 + e^{\beta p^2/{2m} - \eta_q}}  ,   \\
 \eta_q &=& \beta \left(\mu_q - u_q\right)
       = \beta p_{Fq}^2/(2m) = p_{Fq}^2/(2mT) = \ln z_q   ,   \\
 \lambda &=& \sqrt{2\pi\hbar^2/mT}  
\end{eqnarray}
we can write, for $f(\vec r, \vec p) = f(\vec r, p)$,
\begin{eqnarray}
 \rho_q &=& \int d^3 p f_q(\vec r, \vec p)
   = \frac{\gamma}{h^3} \int d^3 p \frac{1}{e^{\beta(p^2-p_{Fq}^2)/(2m)} + 1}
   = \lambda^{-3} \frac{2\gamma}{\sqrt{\pi}} F_{1/2}(\eta_q)   ,  \\
 \epsilon_{Fq} &\equiv& \frac{\hbar^2}{2m} 
         \left(\frac{6\pi^2}{\gamma}\rho_q\right)^{2/3}   ,  \\
 {\cal E}_{Kq} &=& \frac{3}{2} P_{Kq}    
  = \int d^3 p \frac{p^2}{2m} f_q(\vec r, \vec p)
  = \frac{\gamma}{h^3} \int d^3 p \frac{p^2}{2m}
        \frac{1}{e^{\beta(p^2-p_{Fq}^2)/(2m)} + 1}     \nonumber \\
  &=& \frac{4\gamma\hbar^2\sqrt{\pi}}{m}
                 \lambda^{-5} F_{3/2}(\eta_q)   
   = \frac{1}{\beta} \frac{2\gamma}{\sqrt{\pi}}
                 \lambda^{-3} F_{3/2}(\eta_q)   
\end{eqnarray}
Here $\epsilon_{Fq}$ is the chemical potential at absolute zero
or Fermi energy and $p_{Fq}$ is the effective Fermi momentum at $T$.
The particle number $N_q = \int d^3 r \rho(\vec r)$ determines
the effective Fermi momentum $p_{Fq}(\vec r)$ or $\eta_q$ at $T$,
in terms of density $\rho_q(\vec r)$,
\begin{eqnarray}
 \eta_q(\rho_q, T) &=&  \beta(\mu_q - u_q) = \beta\frac{p_{Fq}^2}{2m}
     = F_{1/2}^{-1}(\frac{\sqrt{\pi}}{2\gamma} \lambda^3 \rho_q)
\end{eqnarray}

For multi(two)-component systems with potential energy given
by Eq.(\ref{potene}), for a given $\rho_q$ (or $p_{Fq}$) and $T$,
\begin{eqnarray}
 \mu_q(\rho_q,T) &=& T \eta_q(\rho_q, T)  
         \nonumber \\  & &
   + t_0 (1+\frac{x_0}{2}) \rho
      + \frac{t_3}{12} (1+\frac{x_3}{2}) (\alpha+2) \rho^{\alpha+1}
   - \frac{t_3}{12} (\frac{1}{2}+x_3) \alpha \rho^{\alpha+1}
       \nonumber \\  & &
   - t_0 (\frac{1}{2}+x_0) \rho_q
   + \frac{t_3}{12} (\frac{1}{2}+x_3) (\alpha-1) 2 \rho^\alpha \rho_q
   - \frac{t_3}{12} (\frac{1}{2}+x_3) 2\alpha \rho^{\alpha-1} \rho_q^2
         \nonumber \\  & &
  + C \beta\rho^{\beta-1} \rho_p^2 + 2 C \rho^\beta \rho_p \delta_{q,p}
               + \eta C_s \rho^{\eta-1}    ,       \\
 P(\rho_q,T) &=& \sum_q \frac{2}{3} {\cal E}_{Kq}(\rho_q,T)     \nonumber \\
   & & + \frac{t_0}{2} (1 + \frac{x_0}{2}) \rho^2
   + \frac{t_3}{12} (1 + \frac{x_3}{2})(\alpha+1) \rho^{\alpha+2}
       \nonumber \\  & &
   - \frac{t_0}{2} (\frac{1}{2} + x_0) \sum_q \rho_q^2
   - \frac{t_3}{12} (\frac{1}{2} + x_3) (\alpha+1) \rho^\alpha \sum_q \rho_q^2
         \nonumber \\  & &
   + C (\beta + 1) \rho^\beta \rho_p^2  + C_s (\eta - 1) \rho^\eta   ,    \\
 {\cal E}(\rho_q,T) &=& \sum_q {\cal E}_{Kq}(\rho_q,T)
       \nonumber \\  & &
    + \frac{t_0}{2} (1 + \frac{x_0}{2}) \rho^2
              - \frac{t_0}{2} (\frac{1}{2} + x_0) \sum_q \rho_q^2
         + \frac{t_3}{12} (1 + \frac{x_3}{2}) \rho^{\alpha+2}
           - \frac{t_3}{12} (\frac{1}{2} + x_3) \rho^\alpha \sum_q \rho_q^2
                 \nonumber \\  & &
      + C \rho^\beta \rho_p^2 + C_s \rho^\eta      ,    \\
 T\calS(\rho_q,T) &=& \sum_q \frac{5}{3} {\cal E}_{Kq}(\rho_q,T)
                - \sum_q (\mu_q - u_q) \rho_q     ,   \\
 c_P(\rho_q,T) &=& \frac{dQ/A}{dT}
     = T \left(\frac{\partial\calS/\rho}{\partial T}\right)_P  
         \hspace{0.7cm} {\rm or} \hspace{0.7cm}
 c_V(\rho_q,T) ~=~ T \left(\frac{\partial\calS/\rho}{\partial T}\right)_V
\end{eqnarray}
Once we evaluate $F_{1/2}(\eta)$ and $F_{3/2}(\eta)$,
or more directly $\eta = F_{1/2}^{-1}(\chi)$ and $F_{3/2}(\eta)$,
we can evaluate various thermodynamic quantities
in terms of $\rho_q$ and $T$.

For low temperature and high density limit, $\lambda^3\rho$ large,
i.e., when the average de Broglie thermal wavelength $\lambda$ is larger
than the average interparticle separation $\rho^{-1/3}$, 
we can use nearly degenerate (Fermi gas) approximations \cite{huang} 
for $F_{1/2}$ to obtain
\begin{eqnarray}
 \eta_q(\rho_q, T) &=&  \beta(\mu_q - u_q) = \beta\frac{p_{Fq}^2}{2m}
     = F_{1/2}^{-1}(\frac{\sqrt{\pi}}{2\gamma} \lambda^3 \rho_q)
     = \epsilon_{Fq} \left[ 1
             - \frac{\pi^2}{12} \left(\frac{T}{\epsilon_{Fq}} \right)^2
             + \cdots \right]      \nonumber  \\
    &=& \frac{\hbar^2}{2m}\left(\frac{6\pi^2}{\gamma}\right)^{2/3}
      \left[ \rho_q^{2/3}
       - \frac{\pi^2 m^2}{3\hbar^4} \left(\frac{\gamma}{6\pi^2}\right)^{4/3}
               T^2 \rho_q^{-2/3} + \cdots \right]   ,   \\
 {\cal E}_{Kq}(\rho_q, T) &=& \frac{2\gamma}{\beta\sqrt{\pi}}
                 \lambda^{-3} F_{3/2}(\eta_q)   
    = \frac{3}{5} \rho_q\epsilon_{Fq} \left[ 1
         + \frac{5\pi^2}{12} \left(\frac{T}{\epsilon_{Fq}}\right)^2
             + \cdots \right]      \nonumber  \\
   &=& \frac{3\hbar^2}{10m}\left(\frac{6\pi^2}{\gamma}\right)^{2/3}
      \left[ \rho_q^{5/3} + \frac{5\pi^2 m^2}{3\hbar^4}
               \left(\frac{\gamma}{6\pi^2}\right)^{4/3}
               T^2 \rho_q^{1/3} + \cdots \right]
\end{eqnarray}
In the other limit where $\lambda^3\rho$ is small, 
we have a nearly non-degenerate Fermi gas (classical ideal gas)
and the resulting equations are given by an ideal gas 
in leading order with higher order corrections \cite{huang} as
\begin{eqnarray}
 \eta_q(\rho_q, T) &=& \beta (\mu_q - u_q)
    = \ln\left[\frac{\rho_q\lambda^3}{\gamma}
       \left(1 + \frac{1}{2\sqrt{2}} \frac{\rho_q\lambda^3}{\gamma}
         + \cdots \right) \right]
   \approx \ln\left(\frac{\rho_q\lambda^3}{\gamma}\right)
       + \frac{1}{2\sqrt{2}} \left(\frac{\rho_q\lambda^3}{\gamma}\right) 
                 ,    \label{etaqht}   \\
 {\cal E}_{Kq}(\rho_q, T) &=& \frac{2}{3} P_{Kq}
    = \frac{2}{3} \rho_q T\left[1
      + \frac{1}{2^{5/2}} \frac{\rho_q\lambda^3}{\gamma}
      + \left(\frac{1}{8} - \frac{2}{3^{5/2}}\right)
             \left(\frac{\rho_q\lambda^3}{\gamma}\right)^2
      + \cdots \right]    \label{enqht}  
\end{eqnarray}

For a nuclear system with protons and neutrons with the interaction
given by Eq.(\ref{potene}),
the non-degenerate Fermi gas limit of Eqs.(\ref{etaqht}) and (\ref{enqht})
leads to the following set of equations
\begin{eqnarray}
 \mu_q(\rho,y,T) &=& T \ln\left[ \left(\frac{\lambda^3}{\gamma}\right)
                               \rho_q \right]
    + \frac{T}{2\sqrt{2}} \left(\frac{\lambda^3}{\gamma}\right) \rho_q
           \nonumber \\  & &
   + t_0 (1+\frac{x_0}{2}) \rho
      + \frac{t_3}{12} (1+\frac{x_3}{2}) (\alpha+2) \rho^{\alpha+1}
   - \frac{t_3}{12} (\frac{1}{2}+x_3) \alpha \rho^{\alpha+1}
       \nonumber \\  & &
   - t_0 (\frac{1}{2}+x_0) \rho_q
   + \frac{t_3}{12} (\frac{1}{2}+x_3) (\alpha-1) 2 \rho^\alpha \rho_q
   - \frac{t_3}{12} (\frac{1}{2}+x_3) 2\alpha \rho^{\alpha-1} \rho_q^2
         \nonumber \\  & &
  + C \beta\rho^{\beta-1} \rho_p^2 + 2 C \rho^\beta \rho_p \delta_{q,p}
               + \eta C_s \rho^{\eta-1}
               \nonumber \\
  &=& T \ln\left[ \left(\frac{\lambda^3}{\gamma}\right)
        \left(\frac{\rho}{2} \pm (2y-1)\left(\frac{\rho}{2}\right) \right)
            \right]   \nonumber \\  & &
    + \frac{T}{2\sqrt{2}} \left(\frac{\lambda^3}{\gamma}\right)
       \left[\frac{\rho}{2} \pm (2y-1)\frac{\rho}{2}\right]
    + \frac{3}{4} t_0 \rho
    \mp \left(\frac{1}{2} + x_0\right) t_0 (2y-1)\left(\frac{\rho}{2}\right)
           \nonumber \\  & &
    + \frac{(\alpha+2)}{16} t_3 \rho^{\alpha+1}
    - \frac{1}{6} \left(\frac{1}{2} + x_3\right) t_3
          \left[\alpha (2y-1)^2 \left(\frac{\rho}{2}\right)^2
             \pm (2y-1)\left(\frac{\rho}{2}\right) \rho \right]
                 \rho^{\alpha-1}  
           \nonumber \\  & &
    + \frac{1}{4} C \left[\beta + 2 (1 \pm 1)\right] \rho^{\beta+1}
        + C \left[(\beta + 1 \pm 1) (2y-1) \left(\frac{\rho}{2}\right) \rho
                + \beta (2y-1)^2 \left(\frac{\rho}{2}\right)^2 \right]
            \rho^{\beta-1}
           \nonumber \\  & &
    + \eta C_s \rho^{\eta-1}   ,    \\
 P(\rho,y,T) &=& T\rho
         + \frac{T}{2\sqrt{2}} \left(\frac{\lambda^3}{\gamma}\right)
              \left(\frac{\sum_q \rho_q^2}{2}\right) 
                     \nonumber \\  & &
   + \frac{t_0}{2} (1 + \frac{x_0}{2}) \rho^2
   + \frac{t_3}{12} (1 + \frac{x_3}{2})(\alpha+1) \rho^{\alpha+2}
       \nonumber \\  & &
   - \frac{t_0}{2} (\frac{1}{2} + x_0) \sum_q \rho_q^2
   - \frac{t_3}{12} (\frac{1}{2} + x_3) (\alpha+1) \rho^\alpha \sum_q \rho_q^2
         \nonumber \\  & &
   + C (\beta + 1) \rho^\beta \rho_p^2  + C_s (\eta - 1) \rho^\eta
           \nonumber \\
   &=& T \rho
    + \frac{3}{8} t_0 \rho^2 + \frac{(\alpha+1)}{16} t_3 \rho^{\alpha+2}
    + \frac{T}{2\sqrt{2}} \left(\frac{\lambda^3}{\gamma}\right)
         \left(\frac{\rho}{2}\right)^2
    + \frac{(\beta+1)}{4} C \rho^{\beta+2} + (\eta - 1) C_s \rho^\eta
           \nonumber \\  & &
     - \left[ t_0 \left(\frac{1}{2} + x_0\right)
        + \left(\frac{\alpha+1}{6}\right) t_3 \left(\frac{1}{2} + x_3\right)
             \rho^\alpha
        - \frac{kT}{2\sqrt{2}} \left(\frac{\lambda^3}{\gamma}\right)
        - (\beta+1) C \rho^\beta
       \right] (2y-1)^2 \left(\frac{\rho}{2}\right)^2
           \nonumber \\ & &
       + (\beta + 1) C \rho^{\beta+1} (2y-1) \left(\frac{\rho}{2}\right)
               \label{pryt}   ,   \\
 {\cal E}(\rho,y,T) &=& \frac{3}{2} T \rho
     + \frac{3}{2} \frac{T}{2\sqrt{2}} \left(\frac{\lambda^3}{\gamma}\right)
              \left(\frac{\sum_q \rho_q^2}{2}\right) 
                     \nonumber \\  & &
    + \frac{t_0}{2} (1 + \frac{x_0}{2}) \rho^2
              - \frac{t_0}{2} (\frac{1}{2} + x_0) \sum_q \rho_q^2
         + \frac{t_3}{12} (1 + \frac{x_3}{2}) \rho^{\alpha+2}
           - \frac{t_3}{12} (\frac{1}{2} + x_3) \rho^\alpha \sum_q \rho_q^2
                 \nonumber \\  & &
      + C \rho^\beta \rho_p^2 + C_s \rho^\eta
           \nonumber \\
   &=& \frac{3}{2} T \rho
    + \frac{3}{8} t_0 \rho^2 + \frac{1}{16} t_3 \rho^{\alpha+2}
    + \frac{3}{2} \frac{T}{2\sqrt{2}} \left(\frac{\lambda^3}{\gamma}\right)
       \left(\frac{\rho}{2}\right)^2 + \frac{1}{4} C \rho^{\beta+2}
    + C_s \rho^\eta
           \nonumber \\  & &
     - \left[ t_0 \left(\frac{1}{2} + x_0\right)
        + \left(\frac{1}{6}\right) t_3 \left(\frac{1}{2} + x_3\right)
             \rho^\alpha
      - \frac{3}{2} \frac{kT}{2\sqrt{2}} \left(\frac{\lambda^3}{\gamma}\right)
        - C \rho^\beta
       \right] (2y-1)^2 \left(\frac{\rho}{2}\right)^2
           \nonumber \\ & &
       + C \rho^{\beta+1} (2y-1) \left(\frac{\rho}{2}\right)
               \label{eryt}    ,   \\
 T\calS(\rho,y,T) &=& \frac{5}{2} T \rho
        - T \sum_q \rho_q \ln\left(\frac{\lambda^3}{\gamma} \rho_q\right)
      + \frac{T}{2\sqrt{2}} \left(\frac{\lambda^3}{\gamma}\right)
        \left(\frac{\sum_q \rho_q^2}{4}\right)
           \nonumber \\
   &=& T \rho \left[\frac{5}{2}
        - y \ln\left(\frac{\lambda^3}{\gamma} y\rho\right)
        - (1-y) \ln\left(\frac{\lambda^3}{\gamma} (1-y)\rho\right) \right]
           \nonumber \\ & &
      + \frac{T}{2\sqrt{2}} \left(\frac{\lambda^3}{\gamma}\right)
        \frac{[1 + (2y-1)^2]}{2} \left(\frac{\rho}{2}\right)^2
               \label{tsryt}
\end{eqnarray}
Here, for the proton density ($\rho_p$) and neutron density ($\rho_n$),
we defined,
\begin{eqnarray}
& \rho = \rho_p + \rho_n  ,  \hspace{5ex}
 \rho_3 = \rho_p - \rho_n = (2 y - 1) \rho  , 
    \hspace{5ex} y = \rho_p/\rho  ,
    \nonumber \\
& \rho_p = \frac{1}{2} (\rho + \rho_3) = y \rho  ,   \hspace{5ex}
 \rho_n = \frac{1}{2} (\rho - \rho_3) = (1-y) \rho  ,   \\
& \sum_q \rho_q^2 = \frac{1}{2} (\rho^2 + \rho_3^2)
     = \frac{[1 + (2y-1)^2]}{2} \rho^2 = [1 + 2y(y-1)]\rho^2  ,  \nonumber \\
& \sum_q \rho_q^3 = \frac{1}{4} \rho(\rho^2 + 3\rho_3^2)
     = \frac{[1 + 3(2y-1)^2]}{4} \rho^3 = [1 + 3y(y-1)]\rho^3  \nonumber
\end{eqnarray}
The $\pm$ in $\mu_q$ stands $+$ for $q=$proton and $-$ for neutron.
For $x_3 \ne -1/2$ and $C \ne 0$, the $P(\rho)$ curve for different
vales of $y$ at fixed $T$ may cross at some density $\rho$
which was not seen in Ref.\cite{serot}.

For a constant $T$ and constant $P$,
$\rho$ and $y$ are not independent.
Pressure $P$ of Eq.(\ref{pryt}) is a second order polynomial of
 $(2y-1)\left(\frac{\rho}{2}\right)$,
and thus we have, for the range of $0 \le y \le 1/2$,
\begin{eqnarray}
-1 \le ~ (2y-1) &=& 
    \frac{(\beta+1) C \rho^\beta}
       {\left(\frac{1}{2} + x_0\right) t_0
         + \left(\frac{\alpha+1}{6}\right) \left(\frac{1}{2} + x_3\right)
             t_3 \rho^\alpha - \left(\beta+1\right) C \rho^\beta
       - \frac{kT}{2\sqrt{2}} \left(\frac{\lambda^3}{\gamma}\right) }
                \nonumber  \\  & &  \hspace{-1.5cm}
    \mp \frac{2}{\rho} \left[  
      \left(\frac{(\beta+1) C \rho^\beta}
       {\left(\frac{1}{2} + x_0\right) t_0
         + \left(\frac{\alpha+1}{6}\right) \left(\frac{1}{2} + x_3\right)
             t_3 \rho^\alpha - \left(\beta+1\right) C \rho^\beta
       - \frac{kT}{2\sqrt{2}} \left(\frac{\lambda^3}{\gamma}\right) }
           \right)^2 \left(\frac{\rho}{2}\right)^2
                  \right.   \nonumber  \\  & & \hspace{-1.5cm} \left.
      + \frac{T\rho
          + \frac{3}{8} t_0 \rho^2 + \frac{(\alpha+1)}{16} t_3 \rho^{\alpha+2}
          + \frac{(\beta+1)}{4} C \rho^{\beta+2} + (\eta-1) C_s \rho^\eta
          + \frac{T}{2\sqrt{2}} \left(\frac{\lambda^3}{\gamma}\right) 
            \frac{\rho^2}{4} - P}
        {\left(\frac{1}{2} + x_0\right) t_0
          + \left(\frac{\alpha+1}{6}\right) \left(\frac{1}{2} + x_3\right)
             t_3 \rho^\alpha - \left(\beta+1\right) C \rho^\beta
          - \frac{T}{2\sqrt{2}} \left(\frac{\lambda^3}{\gamma}\right) }
       \right]^{1/2}   
    ~ \le 0      \label{y2m1}
\end{eqnarray}
for a given density $\rho$. (Here $+$ sign is allowed
for the case that the first term is negative.)
Without the Coulomb interaction only the second term in the square root
survives and $y(\rho)$ is a single valued function of $\rho$
in the range of $0 \le y \le 1/2$.
The numerator of the second term in the square
root is $P(\rho,y=1/2,T) - P(\rho,y,T)$, i.e., the negative of the pressure
measured with respect to the pressure for a symmetric nuclear system.
Notice here that $t_0$ is negative and $P(y) \ge P(y=1/2)$
for the potential without the Coulomb interaction (see Eq.(\ref{pryt})).
Since we are considering only $-1 \le (2y-1) \le 0$,
we have conditions of
\begin{eqnarray}
 0 & \le & 
   \frac{T\rho
          + \frac{3}{8} t_0 \rho^2 + \frac{(\alpha+1)}{16} t_3 \rho^{\alpha+2}
          + \frac{(\beta+1)}{4} C \rho^{\beta+2} + (\eta-1) C_s \rho^\eta
          + \frac{T}{2\sqrt{2}} \left(\frac{\lambda^3}{\gamma}\right) 
            \frac{\rho^2}{4} - P}
        {\left(\frac{1}{2} + x_0\right) t_0
          + \left(\frac{\alpha+1}{6}\right) \left(\frac{1}{2} + x_3\right)
             t_3 \rho^\alpha - \left(\beta+1\right) C \rho^\beta
          - \frac{T}{2\sqrt{2}} \left(\frac{\lambda^3}{\gamma}\right) }
               \nonumber  \\
   & \le &
    \left[\frac{2 (\beta+1) C \rho^\beta}
       {\left(\frac{1}{2} + x_0\right) t_0
         + \left(\frac{\alpha+1}{6}\right) \left(\frac{1}{2} + x_3\right)
             t_3 \rho^\alpha - \left(\beta+1\right) C \rho^\beta
       - \frac{T}{2\sqrt{2}} \left(\frac{\lambda^3}{\gamma}\right) }
    + 1\right] \left(\frac{\rho}{2}\right)^2
\end{eqnarray}
to have $0 \le y \le 1/2$ of which the boundaries are
related to the pressures at $y = 1/2$ and $y = 0$ respectively.
For non-zero $C$, there is a back bending in $y(\rho)$.
This behaivor is related to the choice of opposite sign for the square
root term in Eq.(\ref{y2m1}) and double values of $y$ for a given $\rho$.
To find the point of back bending, we look at the
$\left(\partial\rho/\partial y\right)_{T,P} = 0$ point.
Using Eq.(\ref{y2m1}) we can find the energy and entropy as a function
of pressure $P$, proton fraction $y$, and temperature $T$, i.e.,
${\cal E}(P, y, T)$ and $\calS(P, y, T)$.
From these results we may study an isobaric phase transition.
Since all the phases have the same pressure in the coexistent region,
$\mu_q(P, y, T)$ may be used to find the coexistent region.

At fixed $T$ and $P$, only one of either $\rho$ or $y$ is
the independent variable.
Thus observables such as $P$, ${\cal E}/\rho$, $\calS/\rho$
may have a discontinuity in $T$ or $y$
when $\left(\frac{\partial\rho}{\partial T}\right)_{y,P}$
or $\left(\frac{\partial\rho}{\partial y}\right)_{T,P}$
diverges.
We can study the behavior of thermodynamic quantities at a fixed $P$
using $d P = 0$ from Eq.(\ref{pryt}),
\begin{eqnarray}
 dP &=& \left\{ \left[T \rho
         - \frac{T}{2\sqrt{2}} \left(\frac{\lambda^3}{\gamma}\right)
              \left(\frac{\sum_q \rho_q^2}{4}\right) \right]
                   \right\} \frac{dT}{T}
                     \nonumber \\  & &
    + \left\{ \left[ T \rho
         + \frac{T}{2\sqrt{2}} \left(\frac{\lambda^3}{\gamma}\right)
              \left({\sum_q \rho_q^2}\right) \right]  \right.
                     \nonumber \\  & &
   + \left[ {t_0} (1 + \frac{x_0}{2}) \rho^2
   + \frac{t_3}{12} (1 + \frac{x_3}{2})(\alpha+1) (\alpha+2) \rho^{\alpha+2}
          \right.  \nonumber \\  & &
   - {t_0} (\frac{1}{2} + x_0) \left(\sum_q \rho_q^2\right)
   - \frac{t_3}{12} (\frac{1}{2} + x_3) (\alpha + 1) (\alpha + 2)
            \rho^\alpha \left(\sum_q \rho_q^2\right)
         \nonumber \\  & &         \left. \left.
   + C (\beta + 1) (\beta + 2) \rho^\beta \rho_p^2
   + C_s (\eta - 1) \eta \rho^\eta  \right]
                            \right\} \frac{d\rho}{\rho}
           \nonumber \\   
   &-& \left\{ \left[ t_0 \left(\frac{1}{2} + x_0\right)
        + \left(\frac{\alpha+1}{6}\right) t_3 \left(\frac{1}{2} + x_3\right)
             \rho^\alpha
        - (\beta+1) C \rho^\beta
        - \frac{T}{2\sqrt{2}} \left(\frac{\lambda^3}{\gamma}\right)
       \right] \left(\frac{\rho\rho_3}{4}\right)     \right. 
           \nonumber \\ & &    \left.
       + (\beta + 1) C \rho^{\beta} \left(\frac{\rho}{2}\right)^2
             \right\} 4 dy
                  \nonumber  \\
  &=& \left\{ \left[ \rho
    - \frac{1}{2} \frac{1}{2\sqrt{2}} \left(\frac{\lambda^3}{\gamma}\right)
       \left(\frac{\rho}{2}\right)^2
    - \frac{1}{2} \frac{1}{2\sqrt{2}} \left(\frac{\lambda^3}{\gamma}\right)
       \left(2y-1\right)^2 \left(\frac{\rho}{2}\right)^2
             \right] \right\} dT
           \nonumber \\ & &
    + \left\{
       \left[T + 2 \tilde b_2 \rho + (\alpha+2) \tilde b_3 \rho^{\alpha+1}
    + \frac{T}{2\sqrt{2}} \left(\frac{\lambda^3}{\gamma}\right)
       \left(\frac{\rho}{2}\right) + (\beta+2) \tilde b_C \rho^{\beta+1}
      + \eta(\eta-1) C_s \rho^{\eta-1}    \right]  \right. 
           \nonumber \\  & &
     - \left[ t_0 \left(\frac{1}{2} + x_0\right)
        + \left(\frac{\alpha+2}{2}\right) \left(\frac{\alpha+1}{6}\right)
              t_3 \left(\frac{1}{2} + x_3\right) \rho^\alpha
        - \left(\frac{\beta+2}{2}\right) (\beta+1) C \rho^\beta
        - \frac{T}{2\sqrt{2}} \left(\frac{\lambda^3}{\gamma}\right)
            \right]  \nonumber \\ & &      \left.
      ~~  \times (2y-1)^2 \left(\frac{\rho}{2}\right) ~~  
     + (\beta+2) (\beta + 1) C \rho^{\beta} (2y-1)
            \left(\frac{\rho}{2}\right)  \right\} d\rho
           \nonumber \\   
   &-& \left\{ \left[ t_0 \left(\frac{1}{2} + x_0\right)
        + \left(\frac{\alpha+1}{6}\right) t_3 \left(\frac{1}{2} + x_3\right)
             \rho^\alpha
        - (\beta+1) C \rho^\beta
        - \frac{T}{2\sqrt{2}} \left(\frac{\lambda^3}{\gamma}\right)
       \right] (2y-1) \left(\frac{\rho}{2}\right)^2    \right. 
           \nonumber \\ & &    \left.
       + (\beta + 1) C \rho^{\beta} \left(\frac{\rho}{2}\right)^2
             \right\} 4 dy
\end{eqnarray}
Thus we have, for a fixed $P$ and $y$,
\begin{eqnarray}
 \left(\frac{\partial\rho}{\partial T}\right)_{y,P}
   &=& - \left\{ \left[ T \rho
         - \frac{T}{2\sqrt{2}} \left(\frac{\lambda^3}{\gamma}\right)
              \left(\frac{\sum_q \rho_q^2}{4}\right) \right]
             \right\}   \frac{\rho}{T}     \nonumber \\
   & & \times  \left\{ \left[ T \rho
         + \frac{T}{2\sqrt{2}} \left(\frac{\lambda^3}{\gamma}\right)
              \left({\sum_q \rho_q^2}\right) \right]  \right.
                     \nonumber \\  & &
   + \left[ {t_0} (1 + \frac{x_0}{2}) \rho^2
   + \frac{t_3}{12} (1 + \frac{x_3}{2})(\alpha+1) (\alpha+2) \rho^{\alpha+2}
          \right.  \nonumber \\  & &
   - {t_0} (\frac{1}{2} + x_0) \left(\sum_q \rho_q^2\right)
   - \frac{t_3}{12} (\frac{1}{2} + x_3) (\alpha + 1) (\alpha + 2)
            \rho^\alpha \left(\sum_q \rho_q^2\right)
         \nonumber \\  & &         \left. \left.
   + C (\beta + 1) (\beta + 2) \rho^\beta \rho_p^2
   + C_s (\eta - 1) \eta \rho^\eta  \right]
                            \right\}^{-1}
           \nonumber \\  
   &=& - \left\{ \rho
    - \frac{1}{2} \frac{1}{2\sqrt{2}} \left(\frac{\lambda^3}{\gamma}\right)
       \left(\frac{\rho}{2}\right)^2
    - \frac{1}{2} \frac{1}{2\sqrt{2}} \left(\frac{\lambda^3}{\gamma}\right)
       \left(2y-1\right)^2 \left(\frac{\rho}{2}\right)^2
             \right\}      \nonumber \\
   & & \times  \left\{
      \left[T + 2 \tilde b_2 \rho + (\alpha+2) \tilde b_3 \rho^{\alpha+1}
         + \frac{T}{2\sqrt{2}} \left(\frac{\lambda^3}{\gamma}\right)
           \left(\frac{\rho}{2}\right) + (\beta+2) \tilde b_C \rho^{\beta+1}
         + \eta(\eta-1) C_s \rho^{\eta-1}    \right]   \right.
             \nonumber \\  & &
        - \left[ t_0 \left(\frac{1}{2} + x_0\right)
           + \left(\frac{\alpha+2}{2}\right) \left(\frac{\alpha+1}{6}\right)
               t_3 \left(\frac{1}{2} + x_3\right) \rho^\alpha
          - \left(\frac{\beta+2}{2}\right) (\beta+1) C \rho^\beta
          - \frac{T}{2\sqrt{2}} \left(\frac{\lambda^3}{\gamma}\right)
              \right]  \nonumber \\ & &    \left.
        ~~  \times (2y-1)^2 \left(\frac{\rho}{2}\right)  ~~
        + (\beta+2) (\beta + 1) C \rho^{\beta} (2y-1)
              \left(\frac{\rho}{2}\right)
                    \right\}^{-1}
\end{eqnarray}
Using this we get the specific heat capacity, from Eq.(\ref{tsryt}),
\begin{eqnarray}
 c_P &=& T \left(\frac{\partial \calS/\rho}{\partial T}\right)_{y,P}
                    \nonumber \\
  &=& T \left[\frac{5}{2} + \frac{3}{2}
        - y \ln\left(\frac{\lambda^3}{\gamma} y\rho\right)
        - (1-y) \ln\left(\frac{\lambda^3}{\gamma} (1-y)\rho\right) \right]
    - \frac{1}{2} \frac{T}{2\sqrt{2}} \left(\frac{\lambda^3}{\gamma}\right)
        \frac{[1 + (2y-1)^2]}{2} \left(\frac{\rho}{4}\right)
                 \nonumber \\
  & & + \left\{ -T 
    + \frac{T}{2\sqrt{2}} \left(\frac{\lambda^3}{\gamma}\right)
        \frac{[1 + (2y-1)^2]}{2} \left(\frac{\rho}{4}\right)   \right\} 
        \left(\frac{1}{\rho}\right)
      T \left(\frac{\partial\rho}{\partial T}\right)_{y,P}
                 \nonumber \\
  &=& T \left[4   
        - y \ln\left(\frac{\lambda^3}{\gamma} y\rho\right)
        - (1-y) \ln\left(\frac{\lambda^3}{\gamma} (1-y)\rho\right) \right]
    - \frac{1}{2} \frac{T}{2\sqrt{2}} \left(\frac{\lambda^3}{\gamma}\right)
        \frac{[1 + (2y-1)^2]}{2} \left(\frac{\rho}{4}\right)
                 \nonumber \\
  & & + \left\{ T 
    - \frac{T}{2\sqrt{2}} \left(\frac{\lambda^3}{\gamma}\right)
        \frac{[1 + (2y-1)^2]}{2} \left(\frac{\rho}{4}\right)   \right\}^2 
           \nonumber \\  
   & & \times    \left\{
      \left[T + 2 \tilde b_2 \rho + (\alpha+2) \tilde b_3 \rho^{\alpha+1}
         + \frac{T}{2\sqrt{2}} \left(\frac{\lambda^3}{\gamma}\right)
           \left(\frac{\rho}{2}\right) + (\beta+2) \tilde b_C \rho^{\beta+1}
         + \eta(\eta-1) C_s \rho^{\eta-1}    \right]   \right.
             \nonumber \\  & &
        - \left[ t_0 \left(\frac{1}{2} + x_0\right)
           + \left(\frac{\alpha+2}{2}\right) \left(\frac{\alpha+1}{6}\right)
               t_3 \left(\frac{1}{2} + x_3\right) \rho^\alpha
          - \left(\frac{\beta+2}{2}\right) (\beta+1) C \rho^\beta
          - \frac{T}{2\sqrt{2}} \left(\frac{\lambda^3}{\gamma}\right)
              \right]  \nonumber \\ & &    \left.
        ~~  \times (2y-1)^2 \left(\frac{\rho}{2}\right)  ~~
        + (\beta+2) (\beta + 1) C \rho^{\beta} (2y-1)
              \left(\frac{\rho}{2}\right)
                    \right\}^{-1}
           \nonumber \\  
  &=& \left[4 T \rho     
        - T \sum_q \rho_q \ln\left(\frac{\lambda^3}{\gamma} \rho_q\right)
      - \frac{1}{2} \frac{T}{2\sqrt{2}} \left(\frac{\lambda^3}{\gamma}\right)
        \left(\frac{\sum_q \rho_q^2}{4}\right) \right] 
             \left(\frac{1}{\rho}\right)             \nonumber \\  & &
     + \left\{ \left[ T \rho
         - \frac{T}{2\sqrt{2}} \left(\frac{\lambda^3}{\gamma}\right)
              \left(\frac{\sum_q \rho_q^2}{4}\right) \right]  \right\}^2
            \left(\frac{1}{\rho}\right)         \nonumber \\  & &
     \times  \left\{ \left[ kT \rho
         + \frac{T}{2\sqrt{2}} \left(\frac{\lambda^3}{\gamma}\right)
              \left({\sum_q \rho_q^2}\right) \right]  \right.
   + \left[ {t_0} (1 + \frac{x_0}{2}) \rho^2
   - {t_0} (\frac{1}{2} + x_0) \left(\sum_q \rho_q^2\right)
          \right.  \nonumber \\  & &
   + \frac{t_3}{12} (1 + \frac{x_3}{2})(\alpha+1) (\alpha+2) \rho^{\alpha+2}
   - \frac{t_3}{12} (\frac{1}{2} + x_3) (\alpha + 1) (\alpha + 2)
            \rho^\alpha \left(\sum_q \rho_q^2\right)
         \nonumber \\  & &         \left. \left.
   + C (\beta + 1) (\beta + 2) \rho^\beta \rho_p^2
   + C_s (\eta - 1) \eta \rho^\eta  \right]
                            \right\}^{-1}
             ,    \\
 c_V &=& T \left(\frac{\partial \calS/\rho}{\partial T}\right)_{y,V}
                    \nonumber \\
  &=& T \left[4 - y \ln\left(\frac{\lambda^3}{\gamma} y\rho\right)
        - (1-y) \ln\left(\frac{\lambda^3}{\gamma} (1-y)\rho\right) \right]
    - \frac{1}{2} \frac{T}{2\sqrt{2}} \left(\frac{\lambda^3}{\gamma}\right)
        \frac{[1 + (2y-1)^2]}{2} \left(\frac{\rho}{4}\right)
                 \nonumber \\
  &=& \left[4 T \rho     
        - T \sum_q \rho_q \ln\left(\frac{\lambda^3}{\gamma} \rho_q\right)
      - \frac{1}{2} \frac{T}{2\sqrt{2}} \left(\frac{\lambda^3}{\gamma}\right)
        \left(\frac{\sum_q \rho_q^2}{4}\right) \right] 
             \left(\frac{1}{\rho}\right)        
\end{eqnarray}
This shows that there is a divergence of $c_P$ at $T$ with
 $\left(\frac{\partial\rho}{\partial T}\right)_{y,P} = \infty$, i.e.,
 $\left(\frac{\partial T}{\partial\rho}\right)_{y,P} = 0$
showing first order phase transition.
This happens when
 $\left(\frac{\partial y}{\partial\rho}\right)_{T,P} = 0$
which may exist in the coexist region.
The specific heat with fixed volume $c_V$ has no divergence.

\section{Phase transition of a finite nucleus with Coulomb interaction}

\begin{figure}
 \centerline{  \epsfxsize=5in  \epsfbox{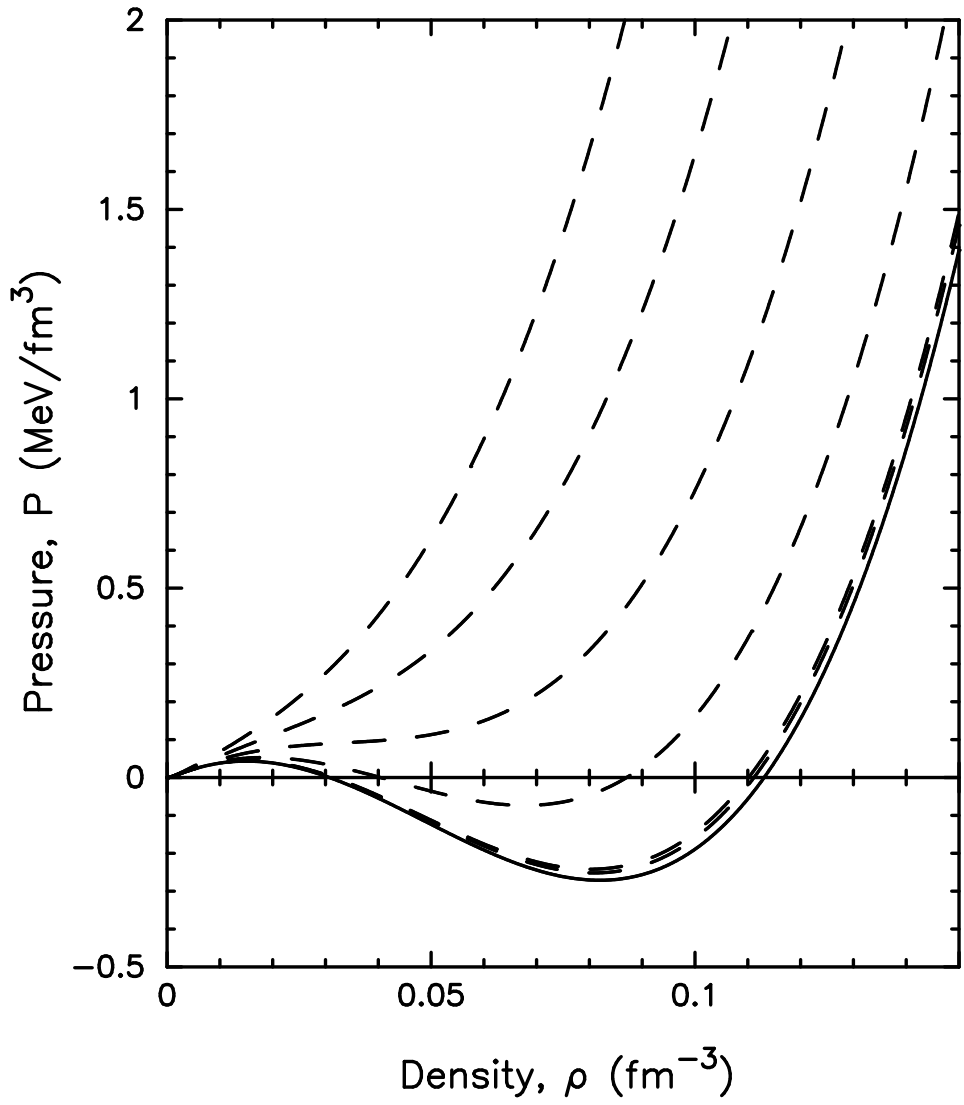}  }
% \centerline{  \epsfxsize=3in  \epsfbox{multfig1.ps}  }
 \vspace{0.5in}
\caption[ ]{   \label{fig1}
Pressure $P(\rho)$ versus $\rho$ at $T = 10$ MeV for various $y$.
The dashed curves from top to bottom
have $y = 0$, 0.1, 0.2, 0.3, 0.4, and 0.5
while the solid curve is for $y = 0.4552$.
  }
\end{figure}

Here we use Skyrme interaction of PRC45 in Table \ref{tabl1}
with $\beta = 0$, $\eta = 2/3$, 
the surface energy $4\pi r_0^2 \sigma = 20.0$ MeV,
and $R = 6$ fm.
Fig.\ref{fig1} shows the pressure $P(\rho)$ for various $y$ values
at $T = 10$ MeV.
With $x_3 = -1/2$, there is no crossing of $P(\rho)$ curves
between different values of $y$ at fixed $T$ which is a behavior
similar to that in Ref.\cite{serot}.
However, for $T = 10$ MeV, the lowest pressure occurs not at $y = 1/2$ but
at $y = 0.4552$ for $R = 6$ fm due to the Coulomb repulsion 
independent of the surface tension (see Eq.(\ref{pryt})).
For $R = 10$ fm, the lowest pressure occurs at $y = 0.3927$.

\begin{figure}
 \centerline{  \epsfxsize=5in  \epsfbox{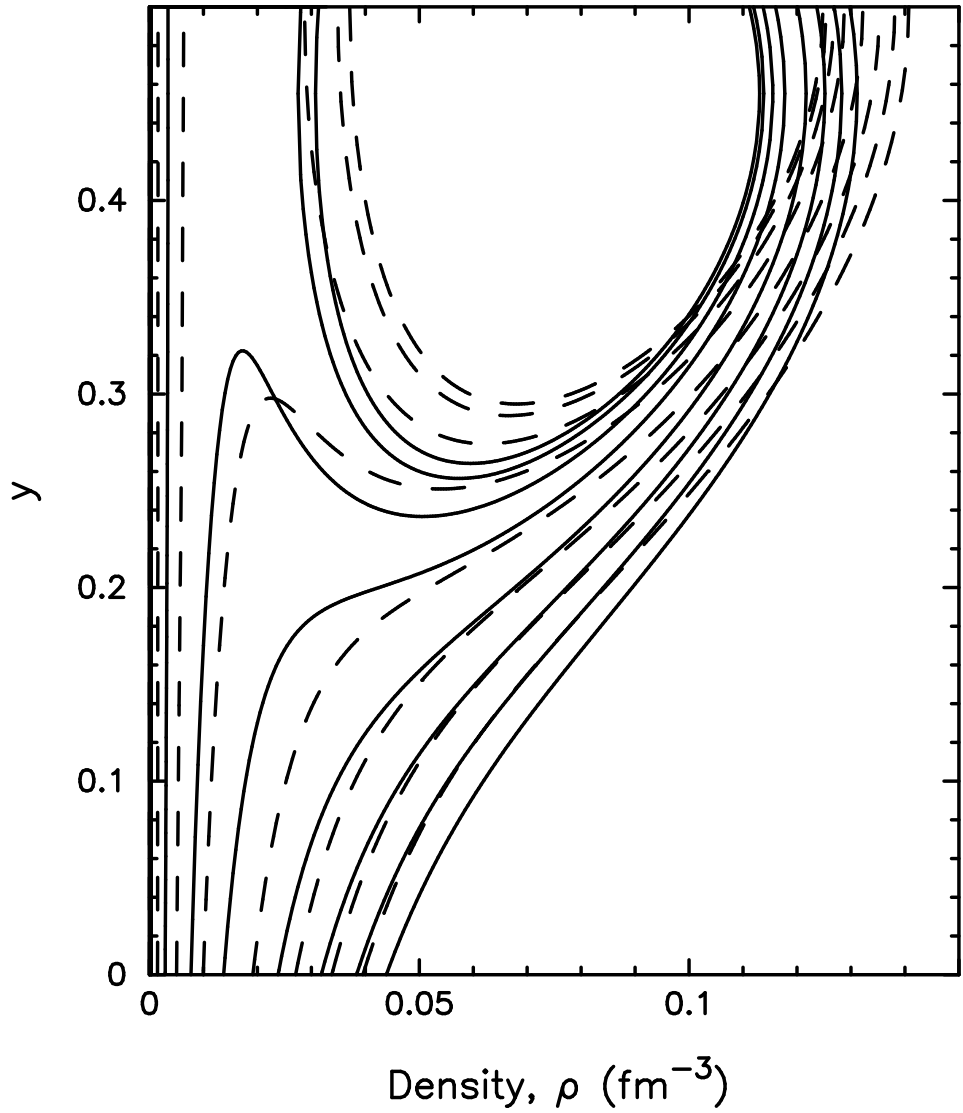}  }
 \vspace{0.5in}
\caption[ ]{   \label{fig2}
Proton fraction
$y(\rho)$ for $P = 0$, 0.015, 0.05, 0.1, 0.2, 0.3, 0.4, and 0.5, from
top to bottom, at $T = 10$ MeV.
The solid curves have Coulomb and surface tension contributions
and the dashed curves are without these interactions.
  }
\end{figure}

For fixed $P$ and $T$, $y$ and $\rho$ are not independent.
Since Eq.(\ref{pryt}) is a second order equation for $y$,
$y$ can have two values for a given density $\rho$ in general.
Fig.\ref{fig2} shows $y(\rho)$ for various $P$ values at $T = 10$ MeV.
Here only the $0 \le y \le 0.5$ region is shown. 
These curves show back bending for large $y$
with $\frac{\partial\rho}{\partial y} = 0$ at $y = 0.4552 < 1/2$
for the parameter with Coulomb interaction which is used here.
At this $y$ value the pressure is minimum as mentioned above.
Without Coulomb interaction, there is no back bending in
the $0 \le y \le 0.5$ region and $\frac{\partial\rho}{\partial y} = 0$
at $y = 1/2$.
The discontinuity of $y(\rho)$ at low density for small pressure is
expected from the low density region of Fig.\ref{fig1}.

\begin{figure}
 \centerline{  \epsfxsize=5in  \epsfbox{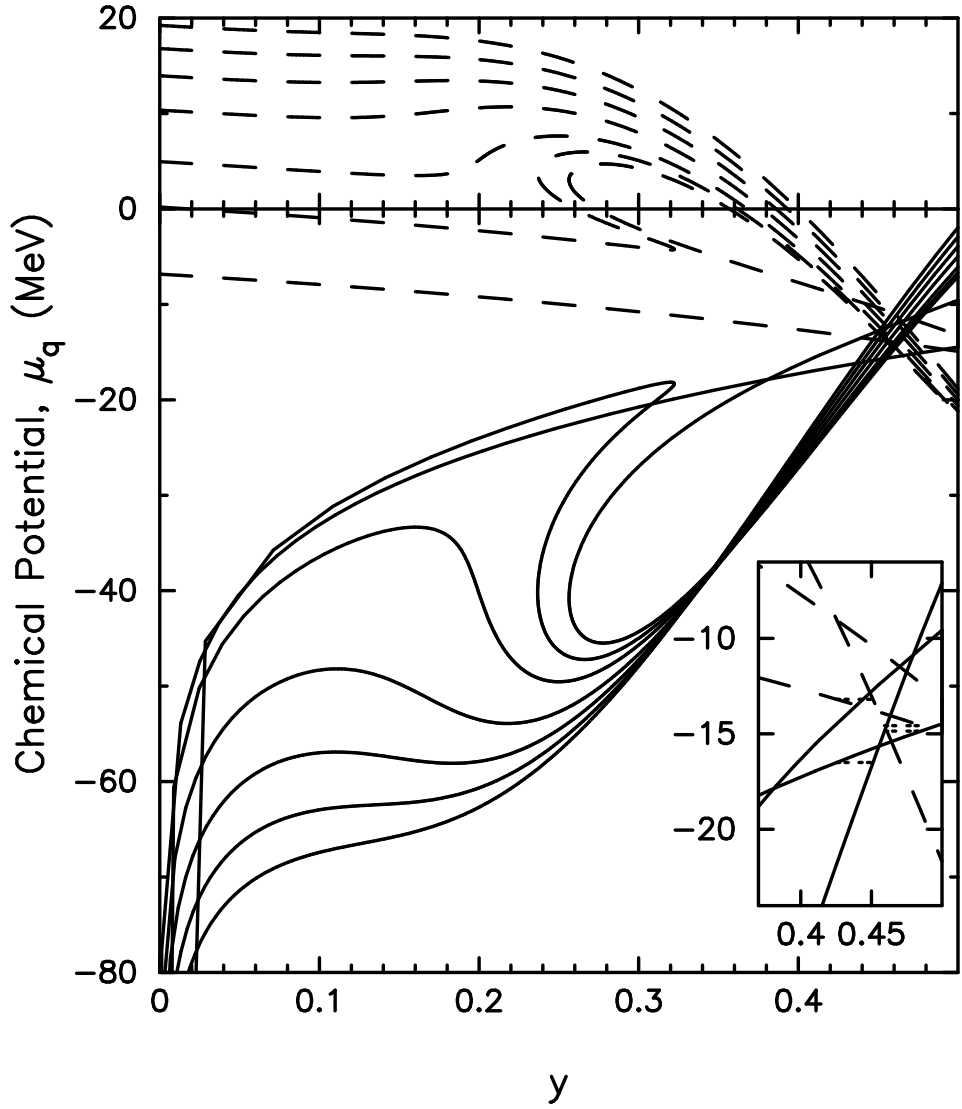}  }
 \vspace{0.5in}
\caption[ ]{   \label{fig3}
Chemical potential $\mu_p(y)$ and $\mu_n(y)$
for $P = 0.015$, 0.05, 0.1, 0.2, 0.3, 0.4, and 0.5,
from top to bottom curve for proton (solid curve) and
bottom to top for neutron (dashed curve) at $T = 10$ MeV.
Chemical potential increases as pressure increase at $y = 1/2$.
$\mu_n$ increases and $\mu_p$ decreases as $P$ increases 
except $\mu_p$ for $P > 0.05$ or 0.1 for smaller $y$.
The small box at the right-bottom corner is the expanded curve
for $P = 0.015$ MeV
with two pairs of binodal points indicated by dotted lines.
  }
\end{figure}

Fig.\ref{fig3} shows chemical potential $\mu_q(y)$ for various $P$
values at $T = 10$ MeV.
From this figure we can find binodal points by checking the 
condition Eq.(\ref{binodmu}).
There is a cross between the small density portion and high
density portion of the curves for low pressure (see the box at the
right-bottom corner of Fig.\ref{fig3}),
which has a discontinuity in $y(\rho)$ shown in Fig.\ref{fig2},
for both $\mu_p$ and $\mu_n$.
The overall behavior is qualitatively similar to the result
of Ref.\cite{serot}.
However with Coulomb interaction $\mu_n < \mu_p$ at $y = 1/2$ 
instead of $\mu_n = \mu_p$.
The existence of these crossings of chemical potential curves
between high and low density and between proton and neutron
allow two pairs of binodal points at low $P$
as shown in the box at the right-upper cornner of Fig.\ref{fig4}.

Fig.\ref{fig4} shows the coexistance binodal curve in $y$-$P$ plane
at $T = 10$ MeV.
The figure shows that the major effect of the surface tension is a
lowering of the pressure for the coexistence curve (compare dash-dotted
curve {\it vs} dashed curve and solid curve {\it vs} dash-dot-dot-dotted
curve).
Inclusion of surface tension may allow for a zero pressure isobaric
phase transition and may simulate the situation of an equilibrated
state of multifragmentation having zero internal pressure of
stable finite nuclei with non-zero gas pressure
as discussed in Ref.\cite{prc56}.
The effect of the Coulomb interaction makes the coexistence curve smaller.
However the more important effect of Coulomb interaction is 
the appearence of another pair of binodal points in the low pressure region
(see the expanded figure in the box at the right upper corner).
Two points of smaller $y$ values have the same chemical potential
and two points of larger $y$ values have the same chemical potential
but which are different (having increased $\mu_p$ and decreased $\mu_n$)
from the other two (see Table \ref{tabl2}).
Here the points with largest and smallest $y$ value are in gas phase
and the points between the other two $y$ values are in the liquid phase
(see Table \ref{tabl2}).
The binodal pair with higher $y$ values allows the mixture of gas
with highest $y$ and liquid with the next high $y$
and is a reversed situation from the other binodal pairs
and the result of Ref.\cite{serot}. 
This allows for a phase separation into a liquid phase with less proton
(lower $y$ value) and a gas phase with more proton (higher $y$ value)
which is more realistic in heavy ion collision.
Two pairs of binodal points meet together with the lowest pressure
of the binodal curve at one point with $y = 0.4552$ for $R = 6$ fm
which would be changed to $y = 0.3927$ for $R = 10$ fm.
Without Coulomb interaction, the lowest pressure of the binodal curve
occurs at $y = 1/2$.

Fig.\ref{fig5} shows $T(\rho)$ for various $P$ values at $y = 0.3$ and 0.5.
This figure shows that at a low pressure there are two points with
$\frac{\partial T}{\partial\rho} = 0$ which make thermodynamic
quantities have a first order discontinulty with $T$ variations.

Fig.\ref{fig6} and Fig.\ref{fig7} 
shows ${\cal E}(T)/\rho$ and $\calS(T)/\rho$
for various $P$ values at $y = 0.3$ and 0.5.
Comparing with the exact result (dotted curve in Fig.\ref{fig6}),
the Fermi gas approximation used here is good for the region
which corresponds to ${\cal E}/\rho > 8$ MeV
for the case without Coulomb and surface effects.
The points with diverging slope of ${\cal E}/\rho$ and $\calS/\rho$
in variations with $T$ are
related to the zero slope points in Fig.\ref{fig5}.
If the charge concentration $y$ was kept constant in all phases
then these two diverging points might be outside of the coexist region
and cause a first order phase transition.
However in the actual phase transition of a heavy ion collision
these diverging points belong to the coexistence region and
gives a second order phase transition as in Ref.\cite{serot}.

%\vspace{0.5in}
%\begin{table}[tbp]
% \label{tabl2}
%\vspace{0.5in}
%\end{table}

\section{Summary and Conclusion}

In this paper we studied the liquid-gas phase transition in
a hadronic system made of protons and neutrons or having two components.
The approach used here was based on a mean filed theory 
and we calculated the mean potential energy using a Skryme interaction.
The work is an extension of a previous study initially begone
in Ref.\cite{skyrmp1} for two components and 
the analysis is similar in some ways to that developed in Ref.\cite{serot}.
However, in this paper we included surface and Coulomb effects
so our results differ quantitavely from those in Ref.\cite{serot}.
Including Coulomb and surface effects introduces some feature also
not present in Ref.\cite{serot}

The importance of surface energies in the liquid-gas phase
transition was shown in Ref.\cite{hirsch,csern}
using an approach based on a statistical model of multifragmentation.
In the statistical model of multifragmentation,
because nuclear matter can be broken into a large number of small pieces,
the surface energy plays a very dominant role.
Here, in a mean field theory, the surface energy plays a lesser
role since we do not allow for the possibility of multifragmentation.
Rather, the system is treated as uniform but with varying density.
Surface tension here brings the coexistance binodal surface to lower
pressure which could allow an isobaric transition at zero pressure.

\begin{figure}
 \centerline{  \epsfxsize=5in  \epsfbox{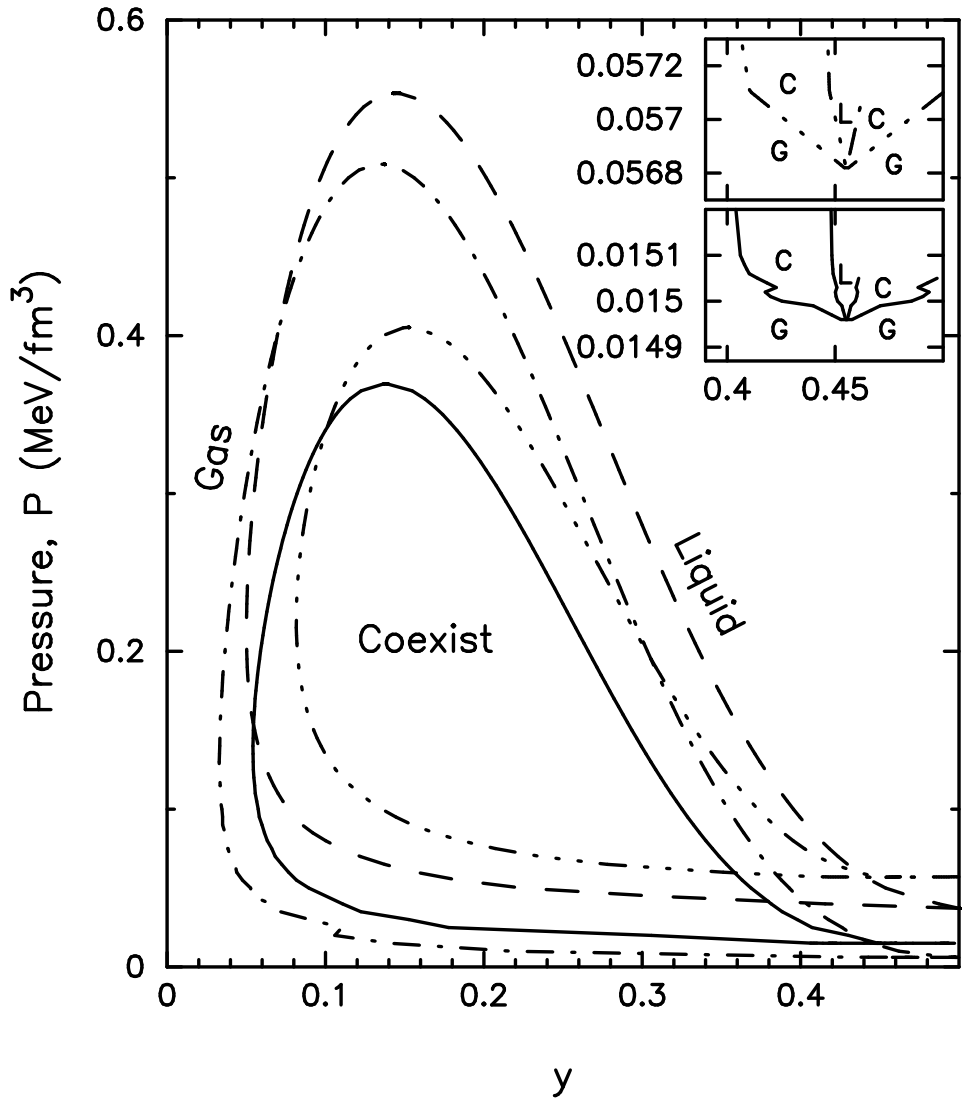}  }
 \vspace{0.5in}
\caption[ ]{   \label{fig4}
Binodal curve at $T = 10$ MeV. 
The solid curve is for the case with Coulomb and surface effect, 
the dash-dotted curve is for the case with Coulomb interaction, 
the dash-dot-dot-dotted curve is for the case with surface effect,
and the dashed curve is for the case without Coulomb and surface effect.
Small boxes at the upper right corner are the expansion of the main figure.
In small boxes, the region of liquid, gas, and coexistence
are indicated by L, G, and C respectively.
  }
\end{figure}

\begin{figure}
 \centerline{  \epsfxsize=5in  \epsfbox{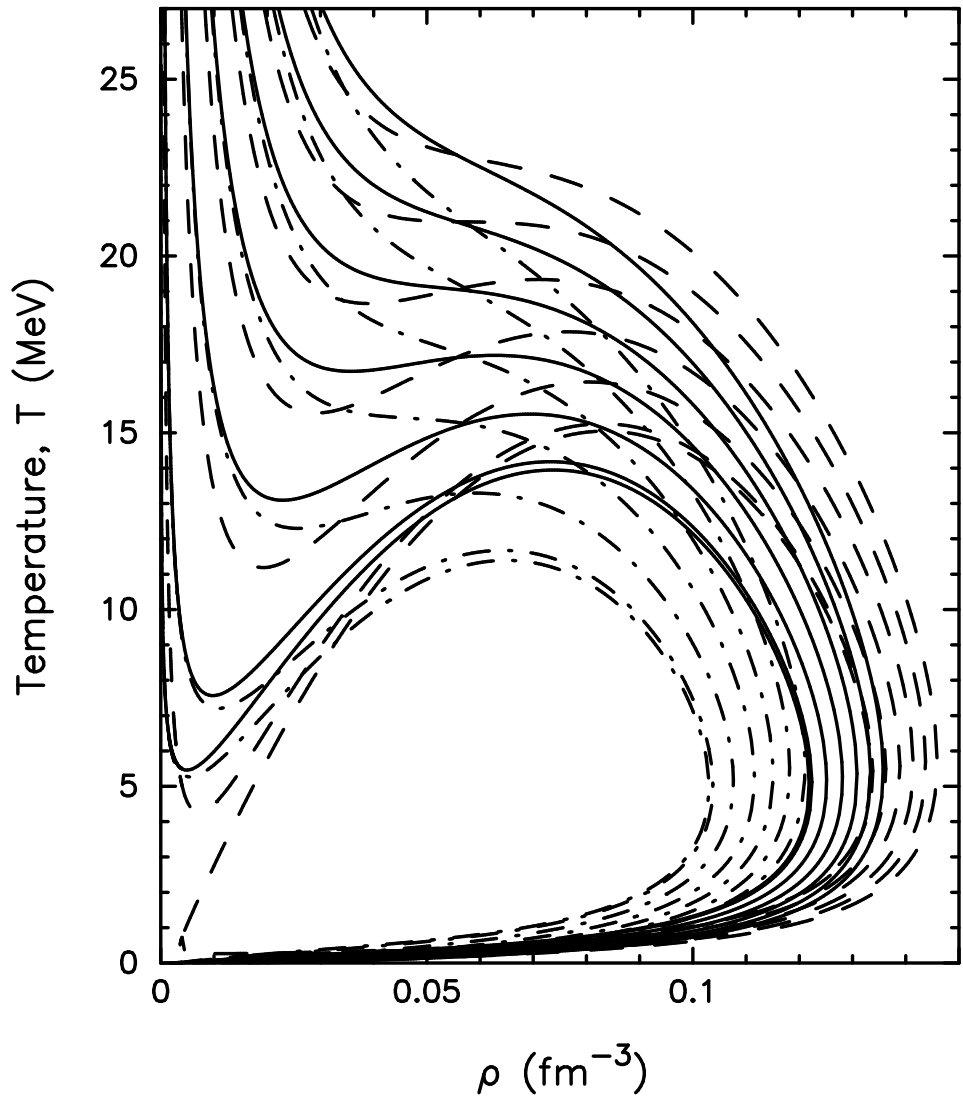}  }
 \vspace{0.5in}
\caption[ ]{   \label{fig5}
Temperature
$T(\rho)$ for $P = 0.0$, 0.015, 0.1, 0.2, 0.3, 0.4, and 0.5, from
bottom to top, at $y = 0.3$ (dash-dotted curve) and 0.5 (solid curve).
Dashed curves are for the case without Coulomb and surface effect.
  }
\end{figure}

\begin{figure}
 \centerline{  \epsfxsize=5in  \epsfbox{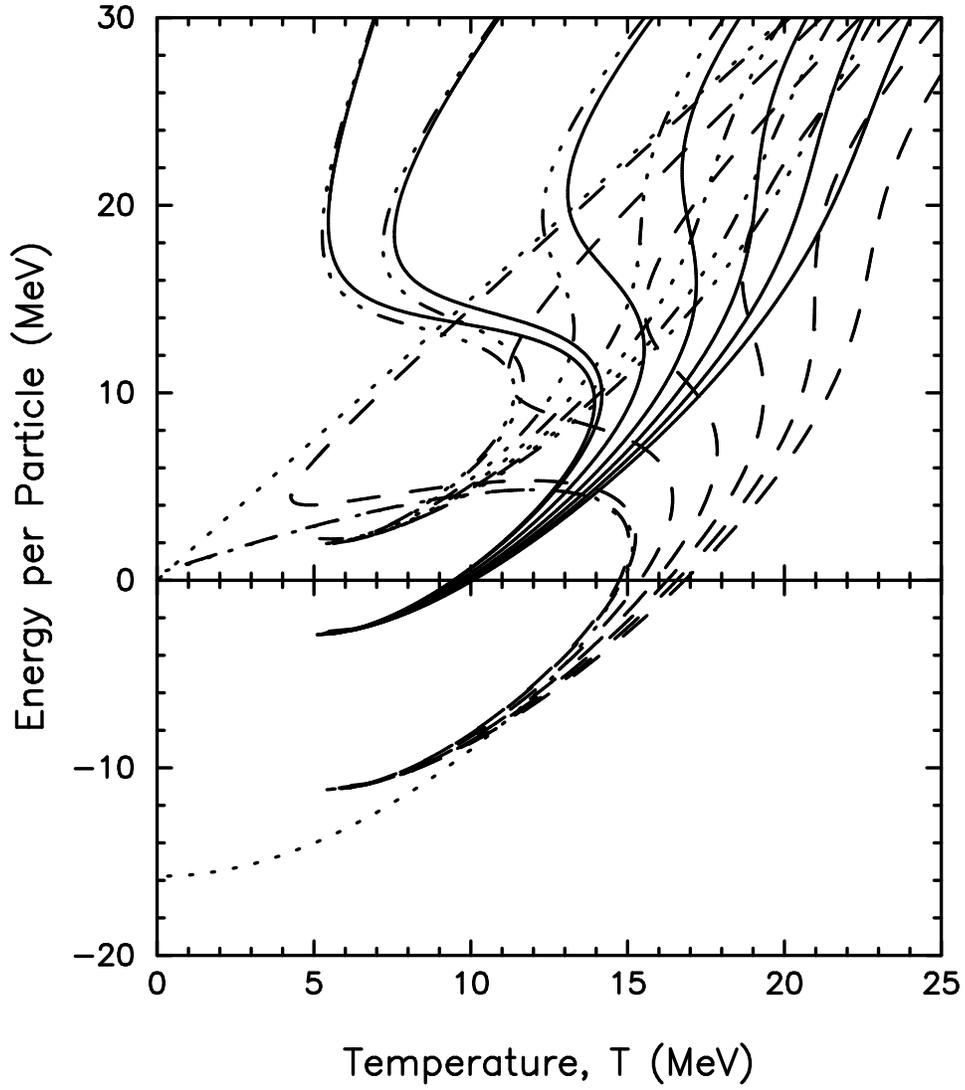}  }
 \vspace{0.5in}
\caption[ ]{   \label{fig6}
Energy per particle ${\cal E}(T)/\rho$ 
for $P = 0.0$, 0.015, 0.1 0.2, 0.3, 0.4, and 0.5, from top to bottom,
at $y = 0.3$ (dash-dot-dot-doted) and 0.5 (solid).
Dashed curves are for $y = 0.5$ without Coulomb and surfase effects and
dotted line is for exact result at $P = 0$ from Ref.\cite{skyrmp2}.
This shows that the Fermi gas approximation we have used is not
varied for $T$ lower than about 5 MeV with high density,
but very accurate for higher temperature (dashed curve for $P = 0.0$
coincide with the dotted curve for ${\cal E}/\rho \ge 8$ MeV).
  }
\end{figure}

\begin{figure}
 \centerline{  \epsfxsize=5in  \epsfbox{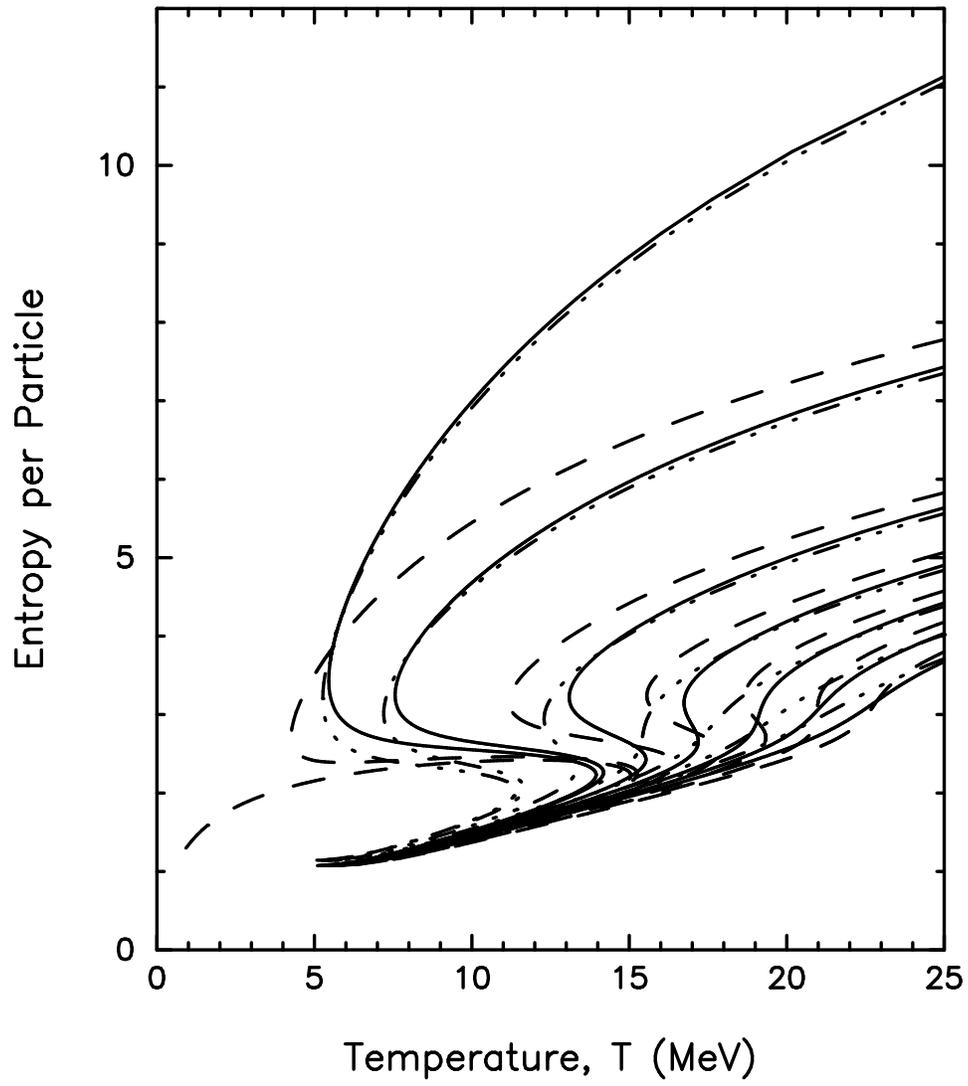}  }
 \vspace{0.5in}
\caption[ ]{   \label{fig7}
Same as Fig.\ref{fig6} but for
entropy per particle $\calS(T)/\rho$ 
  }
\end{figure}

Inclusion of Coulomb interaction brings a new feature
in a mean field approach.
The Coulomb interaction makes the binodal surface smaller
and cause another pair of binodal points at low pressure
with a reversed proton fraction in the liquid-gas phase, i.e.
with less protons in liquid phase amd more protons in gas phase.
The value of the proton fraction $y$ for the liquid phase of
this binodal pair is larger than the $y$ values for the
lowest pressure point of the binodal curve as can be seen in
the box in Fig.\ref{fig4} and we can make 
this value smaller by using a stronger Coulomb interaction.

In this paper we used a fixed value of $R$ to determine
the strength of Coulomb interaction and surface tension.
If we allow for variations of $R$ with $\rho$ or $T$ and
allow different values for the Coulomb and surface energies,
then this simple model may be used to study
multifragmentation with various size of clusters.

%\section{Acknowledgements}

This research was supported by the U.S. DOE, Grant No. DE-FG02-96ER-40987.
S.J. Lee gratefully acknowledges travel support 
for a sabbatical leave from Kyung Hee University
and spent a sabbatical year at Rutgers University in 1999--2000.

\begin{table}[tbh]
\begin{center}
\caption{Parameter sets for Skyrme interaction \protect\cite{skyrmp1,skyrmp2}.}
  \label{tabl1}
\begin{tabular}{cccccc}
 Force & $\alpha$ & $t_0$ (MeV$\cdot$fm$^3$) & $x_0$ &
      $t_3$ (MeV$\cdot$fm$^{3(1+\alpha)}$) & $x_3$ \\
\hline
 \ PRC45 \ &  1  & \ $\frac{4}{3} C_1 = -1089.0$ \ & 1/2 &
          \ \ $\frac{16}{\alpha+2} C_2 = 17480.4$ \ \ & $-1/2$   \\
 ZR1 &  1  & $-1003.9$ & 0.0, 0.2, 0.5 & 13287.2 & 1.0  \\
 ZR2 & 2/3 & $-1192.2$ & 0.0, 0.2, 0.5 & 11041.0 & 1.0 \\
 ZR3 & 0.1 & $-4392.2$ & 0.0, 0.2, 0.5 & 26967.3 & 1.0 \\
\end{tabular}
\end{center}
\end{table}

\begin{table}[tbp]
\begin{center}
\caption{Coexistance points.}
 \label{tabl2}
\begin{tabular}{ccccc}
 $P$ (MeV/fm$^3$) & $\rho$ (fm$^3$) &  $y$  & $\mu_p$ (MeV) & $\mu_n$ (MeV) \\
        \hline 
  0.01496  & 0.0034324 & 0.4526870 & $-15.73360$ & $-13.80920$ \\
           & 0.1138296 & 0.4548638 & $-15.73360$ & $-13.80920$ \\
           & 0.1138296 & 0.4555570 & $-15.59369$ & $-13.92612$ \\
           & 0.0034324 & 0.4577830 & $-15.59369$ & $-13.92612$ \\
  0.01498  & 0.0034352 & 0.4441228 & $-15.97430$ & $-13.61531$ \\
           & 0.1138098 & 0.4536912 & $-15.97430$ & $-13.61531$ \\
           & 0.1138096 & 0.4567391 & $-15.35914$ & $-14.12939$ \\
           & 0.0034352 & 0.4665239 & $-15.35914$ & $-14.12939$ \\
  0.01500  & 0.0034373 & 0.4253351 & $-16.50124$ & $-13.19126$ \\
           & 0.1137651 & 0.4511254 & $-16.50124$ & $-13.19126$ \\
           & 0.1137652 & 0.4592869 & $-14.85386$ & $-14.56769$ \\
           & 0.0034372 & 0.4853333 & $-14.85386$ & $-14.56769$ \\
  0.01503  & 0.0034421 & 0.4226954 & $-16.57709$ & $-13.12985$ \\
           & 0.1137598 & 0.4507550 & $-16.57709$ & $-13.12985$ \\
           & 0.1137592 & 0.4596993 & $-14.77187$ & $-14.63844$ \\
           & 0.0034420 & 0.4883935 & $-14.77187$ & $-14.63844$ \\
  0.01505  & 0.0034449 & 0.4141177 & $-16.81862$ & $-12.93537$ \\
           & 0.1137397 & 0.4495787 & $-16.81862$ & $-12.93537$ \\
           & 0.1137389 & 0.4608853 & $-14.53660$ & $-14.84241$ \\
           & 0.0034448 & 0.4971483 & $-14.53660$ & $-14.84241$ \\
  0.01506  & 0.0034463 & 0.4102457 & $-16.92777$ & $-12.84748$ \\
           & 0.1137306 & 0.4490471 & $-16.92777$ & $-12.84748$ \\
\end{tabular}
\end{center}
\end{table}

\end{document}